\documentclass{article}

\usepackage{arxiv}

\usepackage[utf8]{inputenc} 
\usepackage[T1]{fontenc}    
\usepackage{hyperref}       
\usepackage{url}            
\usepackage{booktabs}       
\usepackage{amsfonts}       
\usepackage{nicefrac}       
\usepackage{microtype}      
\usepackage{lipsum}
\usepackage{graphicx}
\usepackage{subcaption}
\graphicspath{ {./figures/} }

\title{Event Flow - How Events Shaped the Flow of the News, 1950-1995}

\author{
 Melvin Wevers \\
  Department of History\\
  University of Amsterdam, the Netherlands\\
  \texttt{melvin.wevers@uva.nl} \\
   \And
 Jan Kostkan \\
  Center for Humanities Computing Aarhus\\
  Aarhus University, Denmark\\
  \texttt{jan.kostkan@cas.au.dk} \\
  \And
 Kristoffer L. Nielbo\\
  Center for Humanities Computing Aarhus\\
  Aarhus University, Denmark\\
  \texttt{kln@cas.au.dk} \\
}

\begin{document}
\maketitle
\begin{abstract}
This article relies on information-theoretic measures to examine how events impacted the news for the period 1950-1995. Moreover, we present a method for event characterization in (unstructured) textual sources, offering a taxonomy of events based on the different ways they impacted the flow of news information. The results give us a better understanding of the relationship between events and their impact on news sources with varying ideological backgrounds.
\end{abstract}


\section{Introduction}
Huddled around television sets or with ears clung to radio receivers, people all around the world heard Neil Armstrong utter the words: ``[t]hat's one small step for man; one giant leap for mankind.'' This landmark event took place on July 21, 1969. The date denotes the day on which the landing took people, even though one could convincingly argue that this day is part of a sequence of events leading up to this historic moment. In the days before the landing, newspapers published articles that counted down to the event and added commentary to the event, fueling anticipation in public discourse. On a longer time scale, the moon landing was part of a larger event: the space race, a competition between the United States and the Soviet Union for technological dominance. This distinction calls to mind Fernand Braudel's famous description of events as ``surface disturbances, crests of foam that the tides of history carry on their strong backs~\cite{braudel_1995}.''

Events, such as the moon landing, are essential for our experience of history. We do not perceive time as is but through our experience of change in which events demarcate historical temporality. We rely on events to structure the world around us, as individuals and as societies~\cite{wagner-pacificiWhatEvent2017}. William H. Sewell, Jr. describes an event as ``an occurrence that is remarkable in some way - one that is widely noted and commented on by contemporaries~\cite{sewell_historical_1996}.''
In the book \textit{What is an Event?}, Wagner-Pacifici uses 9/11 as a key example to theorize about the form and flow of events. She points out that historians have been preoccupied with bounding events in time and space, while she emphasizes ``the \textit{ongoingness} of events.'' As an event unfolds, it disrupts the historical flow while the public tries to make sense of what is happening. Afterward, the public reflects on these events and sets out to integrate these events into a historical narrative. As events gain traction, they transform how we experience historical time.

Understanding how historical temporality differs from natural temporality is crucial for ``understanding how history has shaped the identity of modern society and culture.\cite{koselleck_futures_2004}'' History is a process of both remembering and forgetting events and their relations. For contemporary, the moon landing was a singular event unlike any other, yet, canonized history knows more than one of these singular events. Was the moon really as impactful at the time, or has the wheel of time strengthen its position in our collective memories.

As Wagner-Pacifici points out, events cannot always be tied to exact dates, even though historical events are often connected to specific dates, such as the moon landing, the fall of the Berlin Wall, or winning a European soccer final. Rather than only departing from specific dates in a top-down manner, can we also detect the unfolding of events and their impact on the historical flow in a more data-driven, bottom-up manner? This paper sets out to answer this question by analyzing the relationship between events and the historical flow, represented by the information presented on the front pages of newspapers.

In our case, we model the ways events impacted language use. More specifically, we examine disruptions in the information flow of news on front pages. For example, events can disrupt the flow of the news by decreasing the amount of novel information presented on the front pages. In the run-up to an event, an increasing focus of the public's eye might be reflected in the increasing uniformity of discourse. Alternatively, an event could have a sudden impact while retaining the public's attention for an extended period. One could hypothesize different archetypical forms of events. In what follows, we try to establish universal motifs, or event flows, from the data itself. The three central questions to this paper are: (1) Do events impact historical flow, as represented by front pages in newspapers? (2) Can we cluster events based on the way they impacted the flow of information? (3) Can we use these clusters to query for events? We call these clusters, event flows, as they represent generalized manners in which events have impacted historical flow.\footnote{Data and code supporting this paper have been made available at \url{https://doi.org/10.5281/zenodo.5509949} (data) and \url{https://github.com/melvinwevers/event-flow} (code).}

A recent special forum in the journal \textit{History and Theory} clearly describes the long-standing historiographical debate on the concept of the event~\cite{jungTimesEventIntroduction2021}. One of the main challenges in history is to combine theoretical work on events with empirical studies of the temporality of events and their relationship to collective memory. The authors claim that systemic analysis of the temporal nature of events, which could shed light on an event's identity, are largely unexplored. This paper offers a computational method that contributes to this effort to understand better how events and their temporal structure have affected public discourse and, by extension, collective memory. 

\section{Related Work}
Previous studies have shown that word usage in newspapers is sensitive to the dynamics of socio-cultural events~\cite{guldi_measures_2019, van_eijnatten_eurocentric_2019, daems_workers_2019}. Furthermore, the co-occurrence of words in newspaper reporting has been shown to capture thematic development accurately~\cite{newman_probabilistic_2006}, and, when modeled dynamically, is indicative of the evolution of cultural values and biases~\cite{van_eijnatten_eurocentric_2019, paul2019bursty, wevers_using_2019}. Methods from complexity science, such as Adaptive Fractal Analysis, have been used to identify distinct domains of newspaper content based on temporal patterns in word use (e.g., advertisements and articles)~\cite{wevers_tracking_2020} and to discriminate between different classes of catastrophic events that display class-specific fractal signatures in, among other things, word usage in newspapers~\cite{gao_culturomics_2012}. 

Several studies have shown that measures of (relative) entropy can detect fundamental conceptual differences between distinct periods~\cite{guldi_measures_2019, degaetano-ortliebUsingRelativeEntropy2018, kestemont_mining_2014}, concurrent ideological movements (e.g. progressive and conservative politics) \cite{barron_individuals_2018, bos_quantifying_2016}, and even, the development of ideational factors (e.g., creative expression) in writing with a serial structure~\cite{murdock_exploration_2015, nielbo_automated_2019, nielbo_curious_2019}. More specifically, a set of methodologically related studies have applied windowed relative entropy to thematic text representations to generate signals that capture information \emph{novelty} as a reliable content difference from the past and \emph{resonance} as the degree to which future information conforms to said novelty~\cite{barron_individuals_2018, murdock_exploration_2015}. Two recent studies have found that successful social media content shows a strong association between novelty and resonance~\cite{nielbo_trend_2021}, and that variation in the novelty-resonance association can predict significant change points in historical data~\cite{vrangbaek_composition_2021}. 

Our paper builds upon this work and will adapt the windowed relative entropy approach to a method that we call Jump Entropy. This method allows us to examine how events have impacted the flow of information in and between newspapers. We compare time series between newspapers and events, using Dynamic Time Warping Barycenter Averaging (DBA)~\cite{petitjean2011global}.

\section{Data}
\begin{figure}
	\centering
		\includegraphics[width=.75\linewidth]{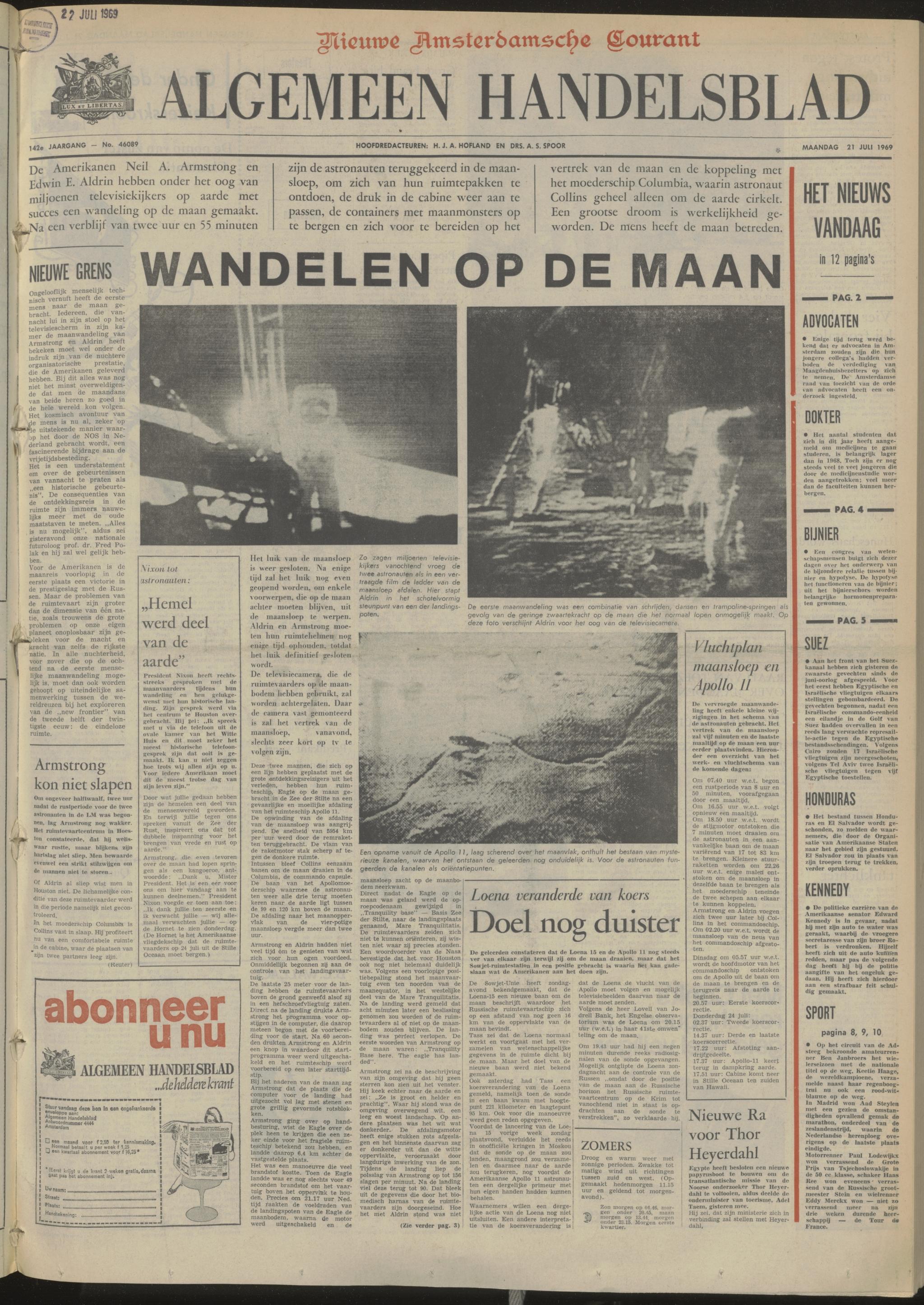}
		\caption{Frontpage of \textit{Algemeen Handelsblad} on July 21, 1969}
	\label{fig:frontpage_example}
\end{figure}

Front pages function as the pulse of the nation, displaying current and pressing events at specific time points. Figure~\ref{fig:frontpage_example}, for example, depicts the front page in \textit{Algemeen Handelsblad} published on the day after the moon landing, which took place on a Sunday. In big, bold letters, we read: ``Walking on the Moon.''\footnote{Translated from the Dutch phrase ``Wandelen op de maan''.} Multiple articles on this event feature on this front page. In addition to the text, we see three images documenting this historic moment. For this study, we only looked into the textual content---captured by optical character recognition (OCR)---and not at the images.
The data consists of the textual content represented on the front pages of ten Dutch national and regional newspapers published between 1950 and 1995 (See Table~\ref{tab:data-overview} for details).\footnote{It is important to note that not all newspapers run for the entire period.} 

\begin{table}[]
    \centering
    \caption{Overview of newspapers in dataset}
    \begin{tabular}{@{}lllll@{}}
        \toprule
            Newspaper & period & type &  \\ \midrule
            Algemeen Handelsblad (AH)           & 1950-1969 &  national &  \\
            De Tijd (DT)                        & 1950-1974 &  national &  \\
            Leeuwarder Courant (LC)             & 1950-1994 &  regional &  \\
            Limburgs Dagblad (LD)               & 1950-1989 &  regional &  \\ 
            Nieuwe Rotterdamsche Courant (NRC)  & 1950-1994 &  national &  \\ 
            Parool (PA)                         & 1950-1995 &  regional &  \\ 
            Telegraaf (TG)                      & 1950-1994 &  national &  \\ 
            Trouw (TR)                          & 1950-1995 &  national &  \\ 
            Volkskrant (VK)                     & 1950-1995 &  national &  \\ 
            Vrije Volk (VV)                     & 1950-1990 &  national &  \\ 
        \bottomrule
    \end{tabular}
    \label{tab:data-overview}
\end{table}

For the data processing, which is not perfect due to flaws in the OCR technology, we removed stop words, punctuation, digits, and words shorter than three and longer than seventeen characters. We lemmatized the text using the NLP toolkit SpaCy.\footnote{https://spacy.io/} Next, we used Latent Dirichlet Allocation (LDA) with collapsed Gibbs sampling to train a topic model of the data.\footnote{Using topic coherence, the optimal number of topics ($k$) centered on 100. Going above or slightly below this number did not impact the results. However, when too few topics are selected the matrix becomes too sparse which makes it difficult to detect shifts in entropy.} 
The input document for topic modeling consisted of a concatenation of all the articles on one single front page. This yields a matrix per newspaper of $P(topic_k|document_d)$ or $\theta$, in this case $document_d$ refers to a front page on a specific date and $topic_k$ holds the probability distribution of topics over documents. These ten matrices functioned as input for the calculation of the Jump Entropy. 

In addition to the newspaper data, we constructed a list of sixty events for 1950-1995, using historical subject-matter knowledge combined with Wikipedia.\footnote{See Appendix \ref{section:Appendix_A} for an overview of these events.} This list includes global and national events. 

\section*{Method}

\paragraph{Jump Entropy} To measure the flow of information between front pages, we propose an adapted version of the approach introduced by~\cite{barron_individuals_2018}. Barron et al.~\cite{barron_individuals_2018} measured the amount of novelty (how unexpected is a document, given previous documents) and transience (the degree to which patterns in documents fade or persist in future documents). They calculate this using varying window sizes, i.e. comparing the novelty of document compared to the average relative entropy contained in a varying number of documents. Relative entropy is a divergence measure that is able to capture the amount of ``surprise'' between two probability distributions, where (in this case) the reader learns to expect one distribution, $\vec{p}$, and then encounters second, say $\vec{q}$. These probability distributions are captured in $\theta$, i.e. the topic distributions from one time point compared to another. In our case, this would be between front pages in one newspaper. 

Calculating novelty and transience using this windowed approach assumes that information accumulates in a continuous flow. This approach is quite sensitive to outliers, especially for shorter time windows. Also, due to the cyclical nature of events (e.g. seasonal or annual events), or the cascading, ripple effect in which an event might have impacted newspaper discourse, taking a continuous window might flatten out these effects. 

To better capture the effect of an event on different time scales and trace ripple effects in public discourse, we adapted their approach. We introduce Jump Entropy, an approach that replaces the shifting window for jumps of different sizes. Rather than moving through the set linearly, we compare sets of front pages that are separated by a given distance. This distance between the two sets is expressed by $J$, the jump size. While using a fixed range of documents (14 days, t - 7 and t + 7), we vary the jump size ($J$) and calculate the JSD between a set of front pages around the focal point $t$ and front pages around a focal point either in the past (negative jump size) or the future (positive jump size).\footnote{We also experimented with shorter time windows, but this adds noise to the signal.} 

While \cite{barron_individuals_2018} compare one front page with a range of front pages, this method compares two ranges of front pages separated by a jump. Put differently, we measure the average entropy for a range of documents and then jump into the past or future and compare this range to a similar range in this period. This approach allows us to measure the amount of ``surprise'' between the focal set to a set in the past or the future; as such, we can spot re-use of themes or recurring debates. Compared to the windowed approach, this method is less sensitive to outliers. We can find cyclical patterns, i.e., which period in the past or future is most similar to the focal period.

In addition to adding jumps, we also used a different metric than~\cite{barron_individuals_2018}. Rather than using Kullback-Leibler (KLD), we used Jensen-Shannon divergence (JSD), a less well-known formulation of relative entropy. JSD has several favorable properties when dealing with cultural information that is not produced in a strictly one-directional fashion. While newspapers are published day by day, the information represented in the papers is not necessarily produced in a one-directional fashion. Articles might have been written earlier, or authors might reflect back onto earlier events. We contend that JSD better reflects these assumptions. First and foremost, JSD is symmetric ensuring that $JSD(P|Q) = JSD(Q|P)$ for probability distributions $P$ and $Q$. Second, as a smooth version of KLD, JSD is well-behaved when $P$ and $Q$ are small. Finally, the square root of JSD is a proper distance metric that can be, for example, be used for clustering probability distributions. A disadvantage of JSD compared to KLD is that it is more computationally costly. However, this additional cost does not significantly impact the current study. 

We model the difference between articles $s^{(j)}$ and $s^{(k)}$ as their relative entropy: 

\begin{equation}
    JSD (s^{(j)} \mid s^{(k)}) =  \frac{1}{2} D (s^{(j)} \mid M) + \frac{1}{2} D (s^{(k)} \mid M)\label{eq:4}
\end{equation}

\noindent with $M = \frac{1}{2} (s^{(j)} + s^{(k)})$ and $D$ is the Kullback-Leibler divergence:

\begin{equation}
    D (s^{(j)} \mid s^{(k)}) = \sum_{i = 1}^{K} s_i^{(j)} \times \log_2 \frac{s_i^{(j)}}{s_i^{(k)}}\label{eq:5}
\end{equation}

We calculated the average relative entropy between a range ($t$) of topic distributions ($s$) at moment $i$ ($s^{i+t}$) and the same range of documents at moment $j$ ($s^{j+t}$). $t$ ranged from -14 to 14, and the jump size ($J$) ranges between -1500 and 1500 with steps of 15:

\begin{equation}
\mathcal{J}_{J}(i) = \frac{1}{w} \sum_{j=1}^{w} D (s^{(i)} \mid s^{(i + vj)})
  \label{eq:jump_entropy}
\end{equation}

\noindent where $v = -1$ for $t < 0$ otherwise $v = 1$, and $D$ is the distance measure (in this case $JSD$), $w$ is a window size and $J$ is the set of jumps of size $w$, and $t$ is the time point (`direction') at which $\mathcal{J}_{J}(i)$ is computed. 

\begin{figure}
    \centering
    \includegraphics[width=\linewidth]{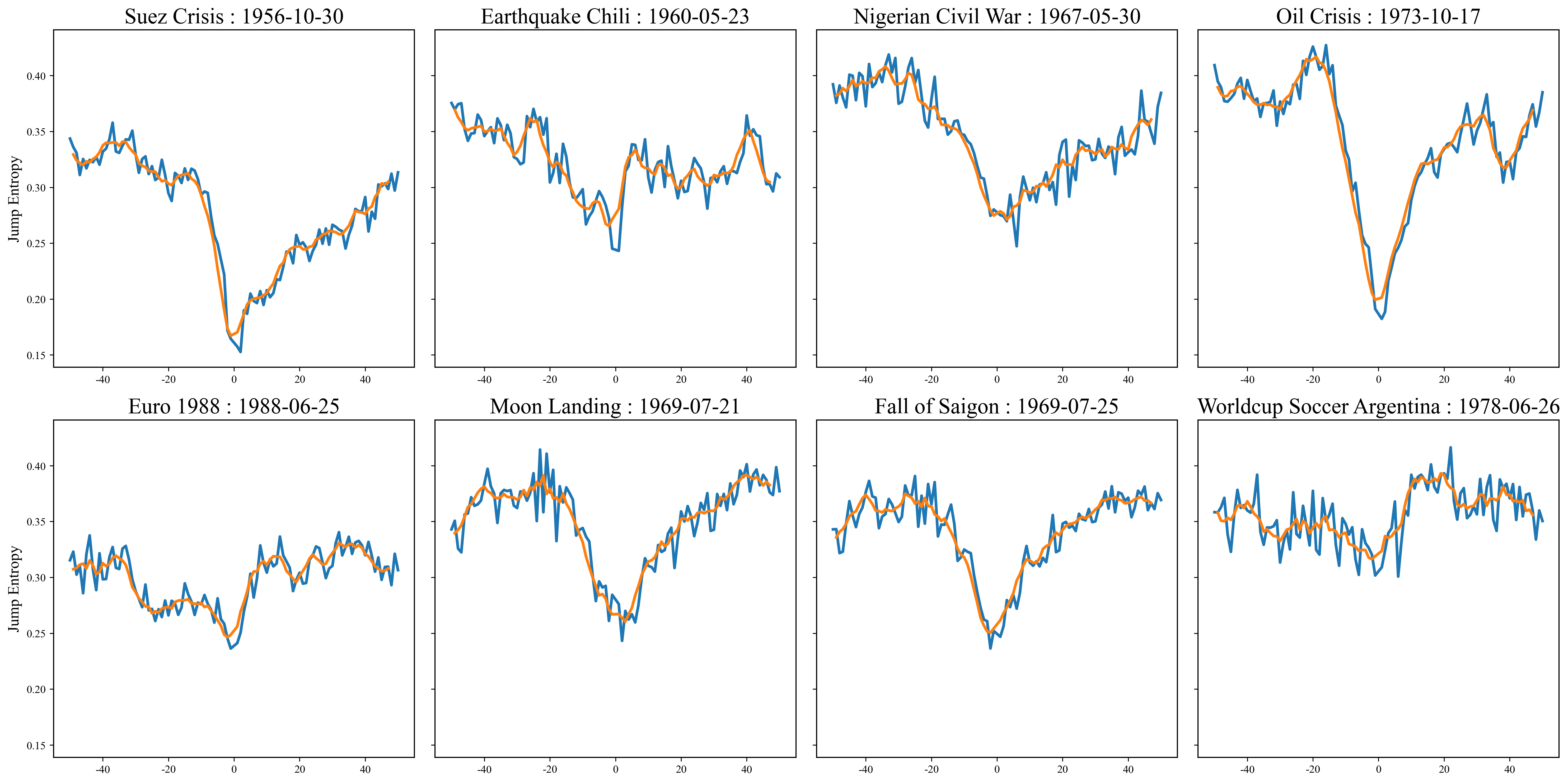}
    \caption{Event Flows of \textit{De Volkskrant}, window size of 50, rolling mean of 5 (orange line).The center of the x-axis reflects the date of the event, and to the left we see jumps in the past, and the right jumps into the future. This graph captures the flow of information leading up to and after the event. On the y-axis, we capture the amount of new information. A lower score means that front pages are more similar, thus a line going down means an increasing focus on a topic.}
    \label{fig:event_flows_vk}
\end{figure}

After calculating the jump entropies for a newspaper, we can use them to visualize event flows. Figure~\ref{fig:event_flows_vk} shows the event flow for eight random event in the newspaper \textit{De Volkskrant}. For each figure, on the x-axis, we see the jump size, and on the y-axis, the relative entropy. The center of the x-axis (0) indicates the date of the event, and to the left we see jumps in the past and to the right jumps into the future. This graph captures the flow of information leading up to and after the event.

\paragraph{Comparing Event Flows}
To group events within and between newspapers in an unsupervised manner requires a method to cluster dynamic processes and compute archetypical (averaged) representations of these time series. Dynamic-Time Warping Barycenter Averaging (DBA) is an ideal solution for exactly that. DBA is based on Dynamic Time Warping (DTW), a technique for optimally aligning time series and flexibly capturing similarities inside the series \cite{petitjean2011global}. As such, DTW accounts for non-linear variations in the time series, i.e., fluctuations do not need to occur at the same time steps~\cite{rakthanmanon2013addressing}. This makes DTW a better distance metric for clustering than traditional Euclidean distance metrics, which have been found to be an inaccurate measure for clustering~\cite{liao2005clustering,petitjean2011global}. 

\begin{figure}
    \centering
    \includegraphics[width=\linewidth]{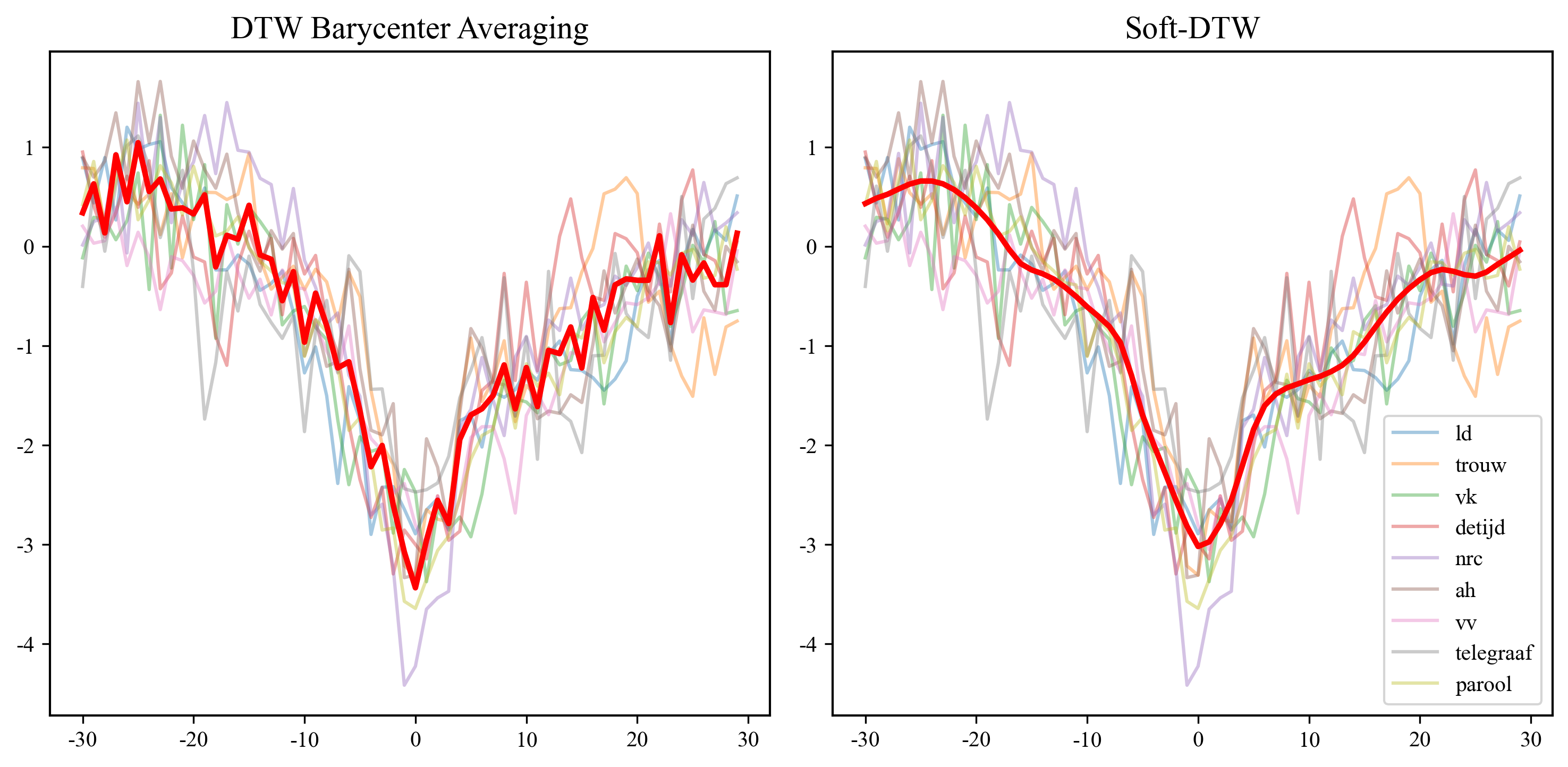}
    \caption{Dynamic Time Warping Barycenter Averaging (DBA) of the time series of the Moon Landing. The red line shows the average time series, while the colored lines show individual time series per newspaper. On the left, we apply the default DBA method, on the right we see Soft-DTW, which uses a differentiable loss function to find the barycenter. The x-axis indicates the window before and after the event (centered at 0).}
    \label{fig:event_flow_moon_landing}
\end{figure}

In principle, DTW allows us to align and compare events between newspapers. However, as pointed out by \cite{petitjean2011global}, while DTW is one of the most used similarity measures for time series, it cannot be reliably used for clustering using well-known algorithms since they rely on K-medoids, which require no averaging. DBA offers an extension of DTW to compute a consensus representation for a set of time series~\cite{petitjean2011global}. This allows us to calculate the average event flow for one event using data from ten newspapers. Figure~\ref{fig:event_flow_moon_landing} gives an example of this process using a DBA and a smoothened version of DTW (soft-DTW) using a soft minimum. \cite{petitjean2011global} show that DBA can be used as input for the k-means clustering of time series. 

\begin{figure}
\centering
\begin{subfigure}{0.5\textwidth}
  \centering
  \includegraphics[width=.9\linewidth]{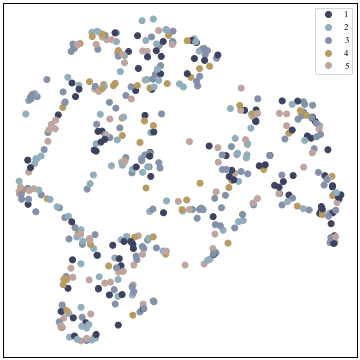}
  \caption{K-means clustering}
  \label{fig:sub1}
\end{subfigure}%
\begin{subfigure}{0.5\textwidth}
  \centering
  \includegraphics[width=.9\linewidth]{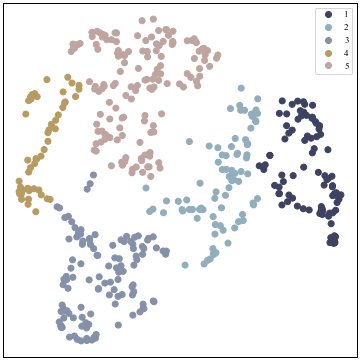}
  \caption{Agglomerative clustering}
  \label{fig:sub2}
\end{subfigure}
\caption{UMAP projection of clusters learned by k-means (left) and agglomerative clustering (right).}
\label{fig:clustering_projections}
\end{figure}

Rather than using k-means clustering, we applied agglomerative clustering. This approach has two main advantages over k-means clustering. First, the method is more explainable; we can inspect how clusters are created, how they are distributed over the dataset, and which clusters are more similar than others. Second, agglomerative clustering led to better separation of the clusters than k-means clustering (see~Figure~\ref{fig:clustering_projections}).

We clustered using the following steps:
\begin{enumerate}
    \item Applying a window size of 28.\footnote{There were four to five clusters for all window sizes between five and fifty. We settled for 28 days for interpretative reasons, as it corresponds to four weeks, or approximately a month of front pages.}
    \item Time series were z-normalized.
    \item Calculate pairwise DTW distance between the events, acquiring a distance matrix.
    \item Project the distance matrix in to two dimensions using UMAP (Uniform Manifold Approximation and Projection).
    \item Grid search through clustering parameters (number of clusters, clustering method), aiming for a high Silhouette score. Additional sanity checks of cluster coherence were taken using the UMAP projection.
    \item After the grid search, euclidean distance was picked as the clustering metric, while UPGMA (unweighted pair group method with arithmetic mean), also known as average linkage, was picked as the linkage criterion. 
    \item Calculate an archetypical time series using DBA for each found cluster.
\end{enumerate}

\noindent Using the described methods, we executed the following steps:

\begin{itemize}
    \item compare similarities and differences between events across newspapers
    \item establish archetypical event flows using agglomerative clustering
    \item use an averaged event flow to query for similar events
\end{itemize}

\section{Results}
In what follows, we will first check whether there exists a difference between newspapers and specific event flows. This step helps us establish for which events there was consensus among newspapers or which newspapers deviated in their reporting on a particular event. Rather than just focusing on our selection of events, we also use a list of random dates as a baseline. 

\begin{figure}
    \centering
    \includegraphics[width=\linewidth]{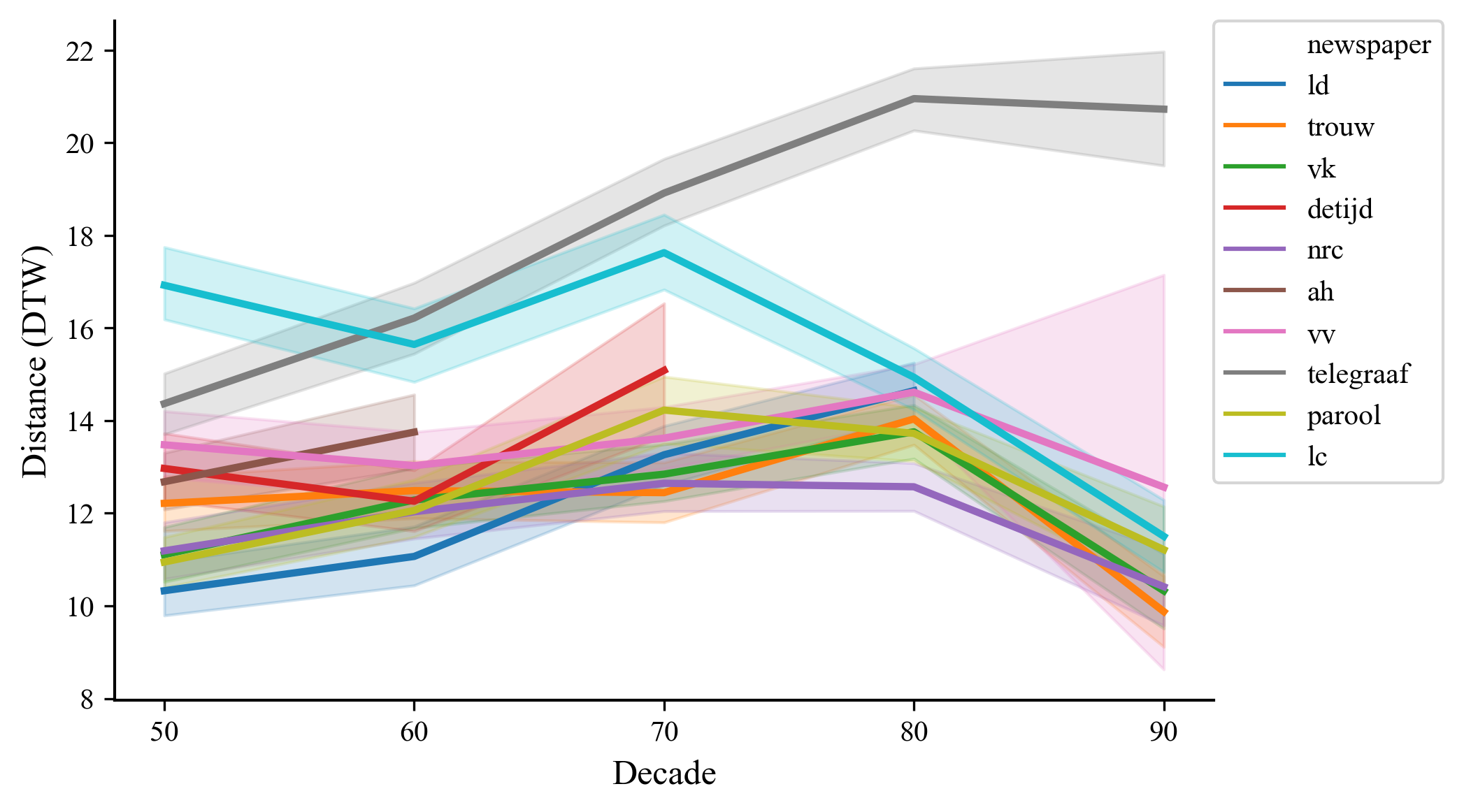}
    \caption{Distance per newspaper to average time series for a random set of 1,000 dates. Mean-aggregated by decade with CI at 95\%.}
    \label{fig:newspaper_difference}
\end{figure}

\paragraph{Newspaper difference using random dates}
We selected the event flows with a jump size of thirty, i.e, thirty days in the future and thirty in the past, for 1,000 random dates between 1950 and 1995 from all the included newspapers. After z-normalizing the time series for every date, we calculated the average event flow per date using DBA. Next, we calculated the distance for each newspaper to each date's average event flow using DTW. This distance to the mean shows us which newspapers deviated the most from the average for that date. In Figure~\ref{fig:newspaper_difference}, we see the distance from the mean per newspaper grouped per decade.

From Figure~\ref{fig:newspaper_difference}, we can gauge that the regional newspaper \textit{Leeuwarder Courant (LC)} and national newspaper \textit{De Telegraaf} deviated the most from the mean, with the latter diverging considerably over the course of these fifty years. This confirms what we knew about the country's most popular newspaper's ideological course, which moved to the right in this period~\cite{hoevenConcentratieKritischeAutonomie2019}. Also, the changing course of the \textit{Leeuwarder Courant} dovetails with the merger of the newspaper with another regional newspaper \textit{Friese Koerier}~\cite{broersmaNieusteTydingenLeeuwarder2002}. It might be that this merger has pushed the newspaper toward to average Dutch newspaper landscape.

\paragraph{Newspaper differences using selected events}
In addition to calculating the difference between papers for random dates, we used our list of events. For each event, we calculated an average event flow using DBA. Next, we calculated the distance from each event per newspaper to the average event flow. From this, we learn for which events the event flows in newspapers were the most similar, and for which events newspapers diverged.\footnote{Since \textit{Algemeen Handelsblad} and \textit{De Tijd} only appeared for a small subset of the period, we excluded these two newspapers}

\begin{figure}
    \centering
    \includegraphics[width=\linewidth]{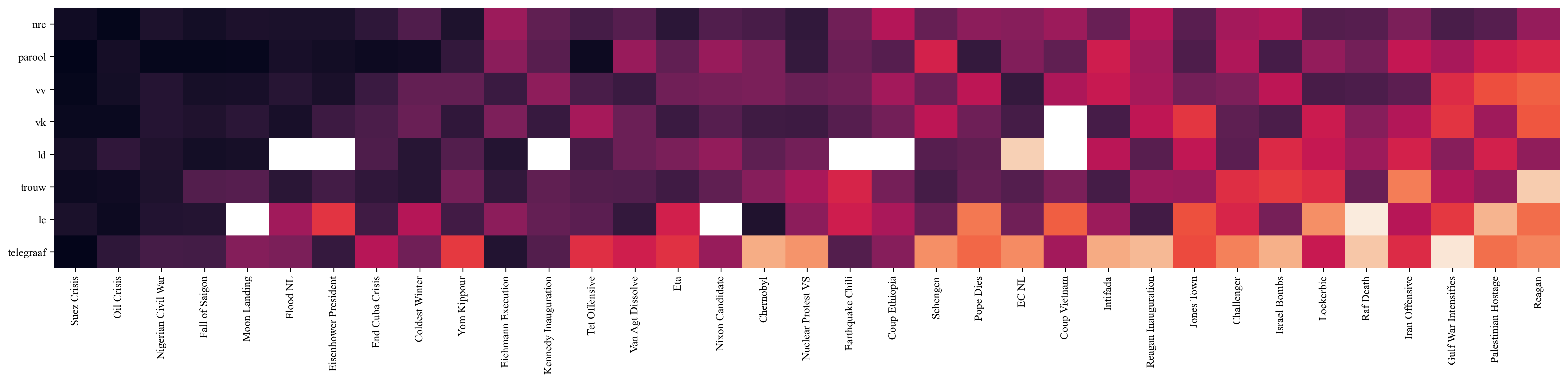}
    \caption{Distance to average event time series. The newspapers and events are sorted based on their mean distance. Dark refers to a shorter distance, while light refers to longer distances. White indicates missing values. For legibiilty, we only included the top 10 events closest to the mean and every other second from the remainder of the list of events.}
    \label{fig:newspaper_event_difference}
\end{figure}

Figure~\ref{fig:newspaper_event_difference} shows that the top five events on which the newspapers reported uniformly were: the Suez crisis in 1956, the 1973 oil crisis, the Nigerian civil war (1967-1970), the fall of Saigon (April 30, 1975), and the moon landing (July 21, 1969). We also see that \textit{NRC Handelsblad}, \textit{Het Parool}, \textit{Het Vrije Volk}, and \textit{De Volkskrant}, were most closely aligned in terms of their event flows, with \textit{De Telegraaf} and \textit{Leeuwarder Courant}, again, being the outliers. Here we also clearly see how \textit{De Telegraaf} is behaving quite distinct compared to the other newspapers. Especially on Middle Eastern affairs, such as the Yom Kippur War (1973) and the Iran hostage crisis (1979-1981), other papers were very much in line with \textit{De Telegraaf} being the exception. 

\begin{figure}
    \centering
    \includegraphics[width=\linewidth]{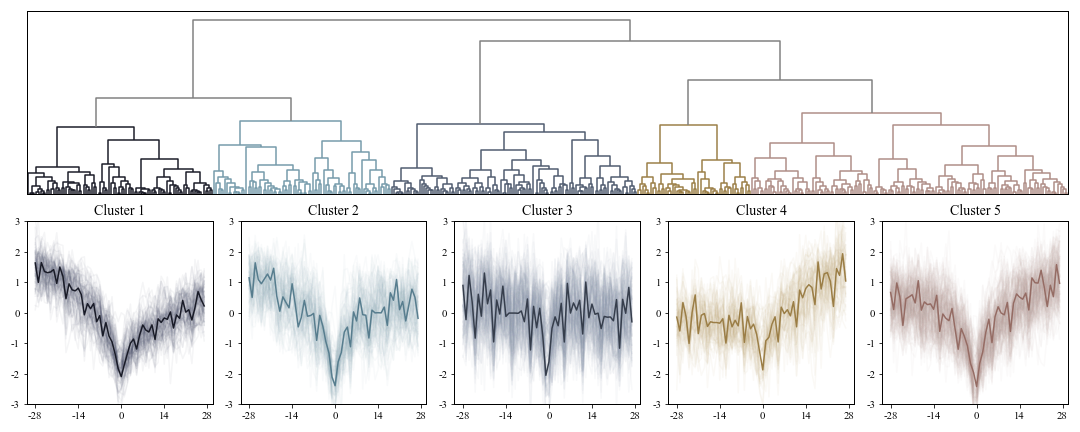}
    \caption{Clustering dendrogram (top) with per-cluster archetypical time series (bottom). DBA time series are indicated by bold lines. Underlying event flows used for the calculation are shown in thin lines. Selected events, window size of 28.}
    \label{fig:event_flow_clusters}
\end{figure}

\paragraph{Archetypical time series}
Using DBA in combination with agglomerative clustering, we looked for clusters of event flows in our data. We excluded \textit{De Telegraaf} because of its deviant behavior. Using a window size of 28, we used the 58 events for the nine remaining newspapers as input. Using Silhouette analysis and cluster separation in the UMAP projection, we determined that the optimal number of clusters was closest to five. Figure~\ref{fig:event_flow_clusters} shows the average event flows within these five clusters. 

From this clustering, we learn that events impacted the news in five characteristic manner. These manner capture how this impact unfolded over time and helps us to understand how events impacted the flow of information in the news, and by extension, how events impacted our historical temporality. The five clusters can be described as follows:

\begin{itemize}
    \item \textbf{Cluster 1}: The downward slope before the event indicates a growing focus on an event, with a slow release indicating persisting, albeit abating focus on the topic after the event.
    \item \textbf{Cluster 2}: The downward slope before the event indicates a growing focus on an event, with a flat line after the event indicative of a persistent focus on a topic after the event. Compared to Cluster 1, the event's impact is more sudden, and it captured the public's attention for a longer period. 
    \item \textbf{Cluster 3}: A noisy pattern that indicates no clear anticipation and a quick release after the event. Events with this signature might have occurred in periods with a quick news cycle, i.e., many news events rapidly superseding each other.
    \item \textbf{Cluster 4}: Stable entropy, indicated by lack of slope, which suggests an increasing focus on a topic in the days before an event. The slope after the event indicates a release of focus after the event. This cluster is the mirror version of Cluster 2 and, to a lesser extent, Cluster 1. 
    \item \textbf{Cluster 5}: This cluster is most similar to cluster 4, albeit more balanced. There is growing anticipation and a release after the event. These event characteristics are indicative of events, such as the Moon Landing, that capture the public's attention in the days before \emph{and} after an event.
\end{itemize}

\paragraph{Querying for Events}
One of the applications of the cluster-based approach is that we can use the average event flow of a cluster (indicated by bold lines), to query for front pages that exhibit a similar pattern. This allows us to search for all the front pages in a particular newspaper that exhibit a sudden focus on a topic, as expressed by Cluster 5. Alternatively, we could also take a specific event, for example, the Oil Crisis in the 1970s, and look for similar events.

\section{Conclusion}
We have presented an adaptation to the method introduced in ~\cite{barron_individuals_2018}, which allows us to capture how events impacted newspaper discourse, and by extension, reveal how the public's eye was drawn to specific events. We have shown how this method can be used to compare how newspapers responded to events and characterize events based on their impact on newspaper discourse. 

The interaction between newspapers and the outside world is a complex interaction. Nonetheless, we managed to characterize ways in which front pages responded to world events. We can use these characterizations to define archetypical time series that can be used to query newspaper data to locate similar events. In this study, we have shown that there were events that impacted the news even though they are not remembered as having an impact, or vice versa. In future work, we will examine how these disjunctions between the public's memory of events and their impact on the news related to the canonization of historical events. 

Also, we found that some noteworthy events displayed no clear signal (cluster 3). For example, the accident with the Challenger space shuttle on January 28, 1986, or the Coup in Ethiopia on December 13, 1960, did not elicit a clear response in the newspapers. For now, we can only speculate about the reasons that these events did not impact the information flow on the front pages of Dutch newspapers. One possibility is that the events did not grasp public attention. Alternatively, it could be that the event was discussed in a more specialized section or that the general public only identified an event as newsworthy well after it occurred.

Closer examination shows that the earthquakes in Chili in May 1960 followed the event flow displayed in Cluster 1, which might seem surprising. However, in this case, there was also a summit with world leaders taking place that increasingly captured the public's attention. The earthquake disrupted this trend and suddenly introduced a new topic, herewith increasing the entropy. 
This example also highlights one of the shortcomings of this approach. Events can overlap each other, move away from the front pages, and after a turn of events they might return to the front again. This movement throughout the papers is not yet captured with this approach. Future work will examine the relationships between topics on the front pages and how they propagated throughout the newspaper. Retention, for instance, could also be expressed by more in-depth reflections on the events in dedicated newspaper sections. 

\section{Acknowledgments}
This study was a NeiC's Nordic Digital Humanities Laboratory project (DeiC-AU1-L-000001), executed on the DeiC Type-1 HPC cluster. We acknowledge The National Library of the Netherlands (KB) for making their newspaper data available. Also, we express our gratitude to Simon DeDeo for his input during the early stages of this paper.

\bibliography{references}

\appendix
\section{Appendix A: Selected Events}
\label{section:Appendix_A}
\begin{tabular}{llll}
\toprule
                event &        date &                           event &        date \\
\midrule
 Eisenhower President &  1953-01-20 &                      Jones Town &  1978-11-18 \\
             Flood NL &  1953-02-02 &                  Snow Storms NL &  1978-12-30 \\
          Suez Crisis &  1956-10-30 &                       Sjah Iran &  1979-01-16 \\
     Earthquake Chili &  1960-05-23 &                      Harrisburg &  1979-03-28 \\
        Coup Ethiopia &  1960-12-15 &                           Salt2 &  1979-06-18 \\
 Kennedy Inauguration &  1961-01-20 &                    Hostage Iran &  1979-11-05 \\
     Startberlin Wall &  1961-08-14 &                   Election Irak &  1980-06-20 \\
   Eichmann Execution &  1962-06-01 &                             Eta &  1980-06-25 \\
      End Cuba Crisis &  1962-10-29 &                   Irak Iran War &  1980-09-22 \\
       Coldest Winter &  1963-01-15 &                 Reagan Election &  1980-11-04 \\
            Pope Dies &  1978-08-07 &             Reagan Inauguration &  1981-01-20 \\
         Coup Vietnam &  1963-11-01 &                 Protest Nuclear &  1981-11-21 \\
           Riot Congo &  1964-11-26 &                   Protest Train &  1982-01-18 \\
      Auschwitz Trial &  1965-08-19 &                    Coup Surinam &  1982-03-11 \\
   Nigerian Civil War &  1967-05-30 &                    Israel Bombs &  1982-04-21 \\
        Tet Offensive &  1968-01-30 &                Van Agt Dissolve &  1982-05-13 \\
            Mlk Death &  1968-04-04 &              Nuclear Protest VS &  1982-06-12 \\
      Nixon Candidate &  1968-08-08 &  Financial Crisis Latin America &  1982-08-12 \\
         Moon Landing &  1969-07-21 &             Heineken Kidnapping &  1983-11-09 \\
       Fall of Saigon &  1969-07-25 &            Gulf War Intensifies &  1984-02-16 \\
           Biafra End &  1970-01-12 &                  Iran Offensive &  1984-10-18 \\
        Bloody Sunday &  1972-01-31 &                       Elfsteden &  1985-02-21 \\
      Olympic Munchen &  1972-09-05 &                 Schengen Accord &  1985-06-14 \\
          Yom Kippour &  1973-10-06 &             Challenger Accident &  1986-01-28 \\
           Oil Crisis &  1973-10-17 &                       Chernobyl &  1986-04-26 \\
  Palestinian Hostage &  1973-12-17 &                        Intifada &  1987-12-09 \\
        Train Hostage &  1975-12-02 &        End Afghan Occupation SU &  1988-04-14 \\
   Crash 747 Tenerife &  1977-03-28 &                    UEFA Eurocup &  1988-06-25 \\
            Raf Death &  1977-10-18 &                       Lockerbie &  1988-12-21 \\
         WC Argentina &  1978-06-26 &                              &          \\
\bottomrule
\end{tabular}

\bibliographystyle{unsrt}  


\end{document}